\documentstyle[aps,prc]{revtex}
\begin{document}
\flushbottom
\widetext
\title
{\bf Reply to Baur's and Bertulani's Comment 
}
\author
{C.J. Benesh}
\address
{Nuclear Theory Center, Indiana University, Bloomington, Indiana, 47405 }
\author
{ A.C. Hayes}
\address
{ AECL, Chalk River
Laboratories, Chalk River, Ontario, Canada K0J 1J0}
\author
{J.L. Friar }
 \address {Los Alamos National Laboratory, Los Alamos, New Mexico, 87455}
\date{\today}
\maketitle
\def\thepage{\arabic{page}}
\makeatletter
\global\@specialpagefalse
\ifnum\c@page=1
\else
\def\@oddhead{\hfill}
\fi
\let\@evenhead\@oddhead
\def\@oddfoot{\reset@font\rm\hfill \thepage \hfill}
\let\@evenfoot\@oddfoot
\makeatother
\begin{abstract}
Shortcomings of a momentum-space treatment of strong absorption, as discussed in the previous Comment,
are only  of concern at low projectile energies, $\gamma < 1.5$.
At intermediate  and high energies,  for which the quantum-mechanical
equivalent-photon spectrum is intended,  the quantum-mechanical  cross sections are reduced relative to the semi-classical results
whether one treats strong effects via a momentum-space or an impact-parameter (spatial) cut-off.  At these energies the origin of the
discrepancy between the predictions of 
fully quantum-mechanical and the  semi-classical calculations   cannot be traced to 
differences in the treatment of strong-interaction effects. Rather, they arise from quantum effects neglected in the semi-classical 
calculations. 
\end{abstract}

\twocolumn
In the previous Comment, the authors contend that 
the discrepancies between the quantum mechanical
 calculations\cite{benesh,beneshfriar} and more traditional 
approaches\cite{bb} to Coulomb excitation in peripheral heavy-ion
collisions are due to an  incomplete treatment of strong absorption in \cite{benesh}. 
They claim that if  strong absorption is treated in terms of a cut-off in the impact
parameter, as opposed to  the cut-off in momentum space used in ref. [1],
then the quantum-mechanical and semi-classical calculations 
would yield the same result.
For low projectile energies, $\gamma < 1.5$, where the cross sections are sensitive to the
details of the treatment of diffractive effects, this criticism is well grounded.
However,   at the intermediate energies of  interest ($1.5 < \gamma < 3$) 
one is already outside
the diffractive limit and strong absorption can be treated adequately in  momentum space.
In this energy range deviations between semi-classical and quantum calculations
arise  mostly from  quantum effects that are normally neglected in calculations of Coulomb excitation.

To disentangle discrepancies  arising from differences in the treatment of strong absorption from discrepancies 
arising from
quantum-mechanical   versus semi-classical treatments of the problem  we make a comparison
between these two approaches and that of an 
an earlier quantum calculation    
of J\"ackle and Pilkuhn\cite{jp}. The advantage in doing this is that J\"ackle and Pilkuhn
calculated 
the Coulomb excitation cross section in the eikonal approximation,
treating strong absorption in  exactly the fashion described 
in the Comment.

The calculations of 
J\"ackle and Pilkuhn find 
cross sections that are  smaller at all energies
than the usual semi-classical results. For all but the lowest projectile energies
($\gamma < 1.5$), the cross sections are reduced  by comparable  amounts  
as we find.
In particular, they find a  reduced photon flux for the mildly relativistic collisions
($1.5 < \gamma < 3.0$) of interest,  and as shown in Fig 2.2 of \cite{bb} this result also holds under the assumption of
pointlike projectiles.

In Fig. 1 we show the equivalent photon number per charge for the two quantum predictions (those   of J\"ackle
and Pilkuhn and of \cite{benesh}), assuming a pointlike projectile, and that 
for the semi-classical predictions \cite{bb} for a 20 MeV  dipole transition.
It is clear
that  at low projectile energies the predictions are quite sensitive to the treatment of diffractive effects
and, as pointed out in the Comment,
the calculation using a momentum-space cut-off deviates considerably from both the semi-classical and the
quantum calculations using a cut-off in the impact parameter. 
However, at intermediate  energies ($\gamma > 1.5$) the predictions of the two quantum calculations,
using a cut-off in coordinate\cite{jp}
versus a cut-off in momentum\cite{benesh} space, are  consistently lower than
the semi-classical calculation. This discrepancy between the quantum and 
semi-classical expressions for the equivalent photon
spectrum cannot be explained
by strong effects. Rather, as noted in \cite{bb}, the effect arises
from quantum and kinematic effects not included in the semi-classical
calculations. 

On a second issue, Baur and Bertulani suggest in their Comment that the inclusion of the  finite size of the projectile does
not affect the predicted cross sections.  On this point we entirely disagree. 
The fact that the projectile  has a form factor 
and that there is a kinematic restriction on 
the magnitude of the three-momentum
transfer in the projectile's rest frame, $\sqrt{-q^2}\ge\omega_T/\gamma\beta$, 
necessarily implies small deviations from  the predictions for a pointlike projectile. 

Finally, Baur's and Bertulani's observation that multi-photon effects are a 
natural consequence of QED is, of course, unassailable. However,
they  
misrepresent our point here. Our claim is that such effects are 
not required to suppress the single-neutron removal cross sections to the levels observed 
experimentally. We stand by that conclusion.

In summary,  the degree of validity of a momentum-space treatment of strong absorption in heavy-ion collisions depends 
 on the projectile-energy range in question.
The concerns expressed in the Comment are justified at low projectile energies ($\gamma < 1.5$),  but not at the energies
of interest for the experimental data under discussion.
At these intermediate energies both momentum-space and coordinate-space treatments of the problem
predict that quantum effects omitted from semi-classical treatments lower the predicted cross sections. There
 appears to be  experimental evidence in support of this  in single-neutron-removal
cross sections, (See Tables III and IV of ref [1]).

\newpage
\begin{figure}

\vspace*{13pt}


\vspace*{2.0truein}             
\includegraphics{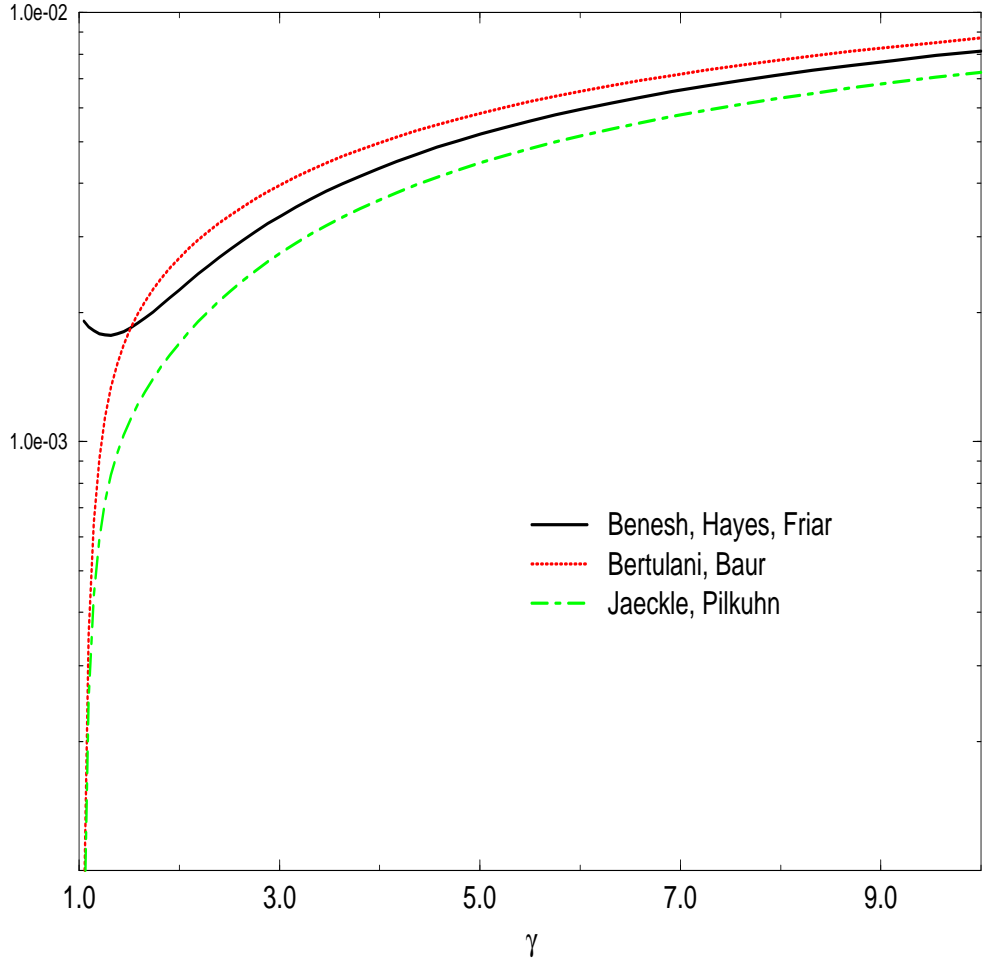}

\vspace*{4in}
\noindent
\caption{ Predictions from the two quantum-mechanical and one semi-classical calculations
for the  equivalent photon number corresponding to a 20 MeV dipole transition.
  The calculation of J\"ackle and Pilkuhn
follows the prescription deemed necessary by Baur and Bertulani  in their Comment.
 Nonetheless, this quantum calculation also finds the  
equivalent photon number to be reduced relative to the semi-classical prediction. Thus, at all but low projectile energies
($\gamma< 1.5$), differences in the treatment of strong absorption effects cannot explain the discrepancies between
the predictions of refs. [1] and [3]. These differences arise from quantum effects omitted from
semi-classical treatments of Coulomb excitation in heavy-ion collisions.}

\end{figure}

\begin{thebibliography}{99}
\bibitem{benesh}C.J. Benesh, A.C. Hayes, and J.L. Friar, Phys. Rev. C{\bf 54} 1404 (1996).\\
\bibitem{beneshfriar}C.J. Benesh and J.L. Friar, Phys. Rev. C{\bf  48}, 1285 (1993).\\
\bibitem{bb} C.A. Bertulani and G. Baur, Phys. Rep. 163, 299 (1988).\\
\bibitem{jp} R. J\"ackle and H. Pilkuhn, Nucl. Phys. {\bf A247},
521(1975).\\
\end{thebibliography}
\end{document}